\documentclass[pdflatex,sn-mathphys-num]{sn-jnl}

\usepackage{graphicx}%
\usepackage{multirow}%
\usepackage{amsmath,amssymb,amsfonts}%
\usepackage{amsthm}%
\usepackage{mathrsfs}%
\usepackage[title]{appendix}%
\usepackage{xcolor}%
\usepackage{textcomp}%
\usepackage{manyfoot}%
\usepackage{booktabs}%
\usepackage{algorithm}%
\usepackage{algorithmicx}%
\usepackage{algpseudocode}%
\usepackage{listings}%

\usepackage{graphicx} 
\graphicspath{{figures/}} 
\usepackage[dvipsnames,svgnames*,x11names*]{xcolor}

\usepackage{geometry}
\usepackage{float}
\usepackage{subcaption}
\usepackage{import}
\usepackage{hyperref}
\usepackage{enumerate}

\newcommand{\textbook}{\textit{textbook}}

\theoremstyle{thmstyleone}%
\theoremstyle{thmstyletwo}%

\theoremstyle{thmstylethree}%

\begin{document}

\title[The search for the gust-wing interaction \textbook{}]{The search for the gust-wing interaction \textbook{}}


\author[1]{\fnm{Paolo} \sur{Olivucci}}\email{p.olivucci@tu-braunschweig.de}

\author*[1]{\fnm{David} \sur{Rival}}\email{david.rival@tu-braunschweig.de}

\affil[1]{\orgdiv{Institute of Fluid Mechanics}, \orgname{TU Braunschweig}, \orgaddress{\street{Hermann-Blenk Str. 37}, \city{ Braunschweig}, \postcode{38108}, \country{Germany}}}

\abstract{
We address whether complex physical relations can be investigated through the synergy of
automated high-volume experiments and the reduction of large datasets to a concise, representative
subset of canonical examples — a \textbook{}. 
To this end, we consider the unsteady aerodynamics of wing-gust interactions, which is
characterized by its rich, high-dimensional physics. 
We take advantage of a purpose-built gust generator to systematically produce over 1,000 distinct
random gust events and to measure the unsteady loads induced on a delta wing.
We then employ a data summarization procedure to identify representative subsets of increasing size
from the large-scale database, which then serve as training data for a machine-learning model of the
aerodynamic loads from sparse pressure measurements. 
An appropriately selected \textbook{} of a few events can achieve predictive accuracy
comparable to random training sets up to two orders of magnitude larger, capturing the intrinsic
diversity of the full-scale data and enhancing modeling efficiency and interpretability.
Our methodology evidences the potential of distilling the essential information contained in large
amounts of 
experimental observations.

}

\maketitle

\section{Introduction}\label{sec:introduction}


The ever-increasing availability of large datasets, computing power and algorithms of diverse scale
and sophistication has enabled, over the last two decades, highly accurate predictive modeling and
pattern recognition in a variety of domains.   
Some authors envision a new paradigm in experimental scientific inquiry, where discoveries are
accelerated by the synergy between the automated collection of vast amounts of raw data and the
application of powerful computational techniques \cite{Fan2019,Angelopoulos2024,Hassabis2024}. 
Instead of traditional hypothesis-testing on a select number of conditions, the collection of a
large database of physical data and its subsequent processing through learning algorithms would
enable accurate predictions and bottom-up identification of explanatory hypotheses. 
More specifically, physics-centric machine learning has found its way into many sub-domains of fluid
mechanics, ranging from ML-aided experimental and computational tools, to turbulence research and
flow control \cite{Brunton2020,DDFM2023,Vinuesa2023}.

In unsteady aerodynamics, resilience to high-intensity atmospheric disturbances is a critical
technical barrier, particularly when transitioning to all-weather autonomous flight at smaller scales \cite{Jones2022}. 
Gust-wing interactions defy comprehensive understanding due to their extreme diversity
including large-amplitude transverse, vortical and streamwise gusts, in either two-dimensional
or fully three-dimensional configurations \cite{Marzanek2019, Jones2021}. 
The high dimensionality of the parameter space encompassed by gust encounters supports the
case for the use of powerful data-driven computer models in the extraction of key gust-wing
interaction physics \cite{Iacobello2022, Kaiser2024}. 
Such models hold promise in effectively uncovering key gust features when special regard is paid to building sparse approaches, aiming at being physically interpretable \cite{Fukami2023}.
Reduced data requirements also lead to leaner predictive models, an aim that is being actively
pursued in data-intensive fields such as natural language processing \cite{Gunasekar2023}, and which is 
relevant to applications in autonomy, where rapid evaluation and less demanding hardware
requirements are decisive factors. 

Recent studies highlight the importance of data-driven approaches to unsteady fluid mechanics, and
informs our effort in building a data-efficient framework for high-dimensional gust-wing
interactions. 
In one instance, a robotic towing tank was designed to autonomously plan and run
experiments on vortex-induced vibrations to map response force coefficients against excitation
parameters \cite{Fan2019}. 
The system employed an active learning algorithm that prioritized exploration of uncertain or
anomalous inputs and could reduce the number of trials needed to confidently model the response to a
fraction of those required by grid search. 
Another study pursued a data-efficient flow reconstruction approach thanks to a neural network model
with embedded physical symmetries, which was able to learn a super-resolution mapping from a single
snapshot of turbulent flow \cite{Fukami2024a}. 
Deep neural networks have also been applied to compressing 
high-dimensional observations
of simulated gusts interacting with a wing section down to a low-dimensional representation, exposing
physically meaningful quantities and enabling high-fidelity flow reconstruction \cite{Fukami2023}.  
A recent experimental investigation builds on a machine-learning aerodynamic load estimation
model to devise an intelligent sparse sensing strategy for a nonslender delta wing impacted by gusts
generated in a towing-tank facility \cite{Chen2023}. 
The success of these efforts promotes further research into data-driven sparsity-promoting
techniques as a valuable tool to navigate the complexity of unsteady flow physics.

Our contribution focuses on a method for discovering and characterizing salient gusts that represent
a minimal collection of representative cases for aerodynamic load prediction algorithms -- a
\emph{textbook} of gust-wing interactions. 
In contrast with other research on data-driven gust physics, our  approach aims at compressing an
initially large database compiled from automated experiments into a maximally informative selection
of examples (data sparsity), rather than reducing its dimensionality or using data to design optimal
experiments.  
The general problem setting is depicted in Figure~\ref{fig:schematic_textbook}a, where an aircraft,
having observed the consequences of number of gusts, is confronted with forecasting the effect of an
incoming disturbance.
Figure~\ref{fig:schematic_textbook}b describes the concept of a \textbook{}, a small summary that
contains the essential information on the physics and enables predictions as accurate as an
extensive database of observations. 

In this article we lay out an approach to search for a \textbook{} as follows.
Section~\ref{sec:problem_statement} formalizes the problem and the notion of \textbook{}.
Section~\ref{sec:experimental_database} illustrates the acquisition of the experimental database and
the aerodynamic load prediction models. 
Section~\ref{sec:learning_curves} weighs the impact that the amount and quality of data have on
predictions and Section~\ref{sec:textbook_selection} presents the \textbook{} selection process.
Section~\ref{sec:discussion} discusses the outcomes. 
The Supplementary Materials contain more technical details on the procedures for data preparation
and textbook selection.

\section{Learning gust-wing interactions from data} \label{sec:problem_statement}

\subsection{Problem description}
\begin{figure*}
\centering
\includegraphics[width=\linewidth]{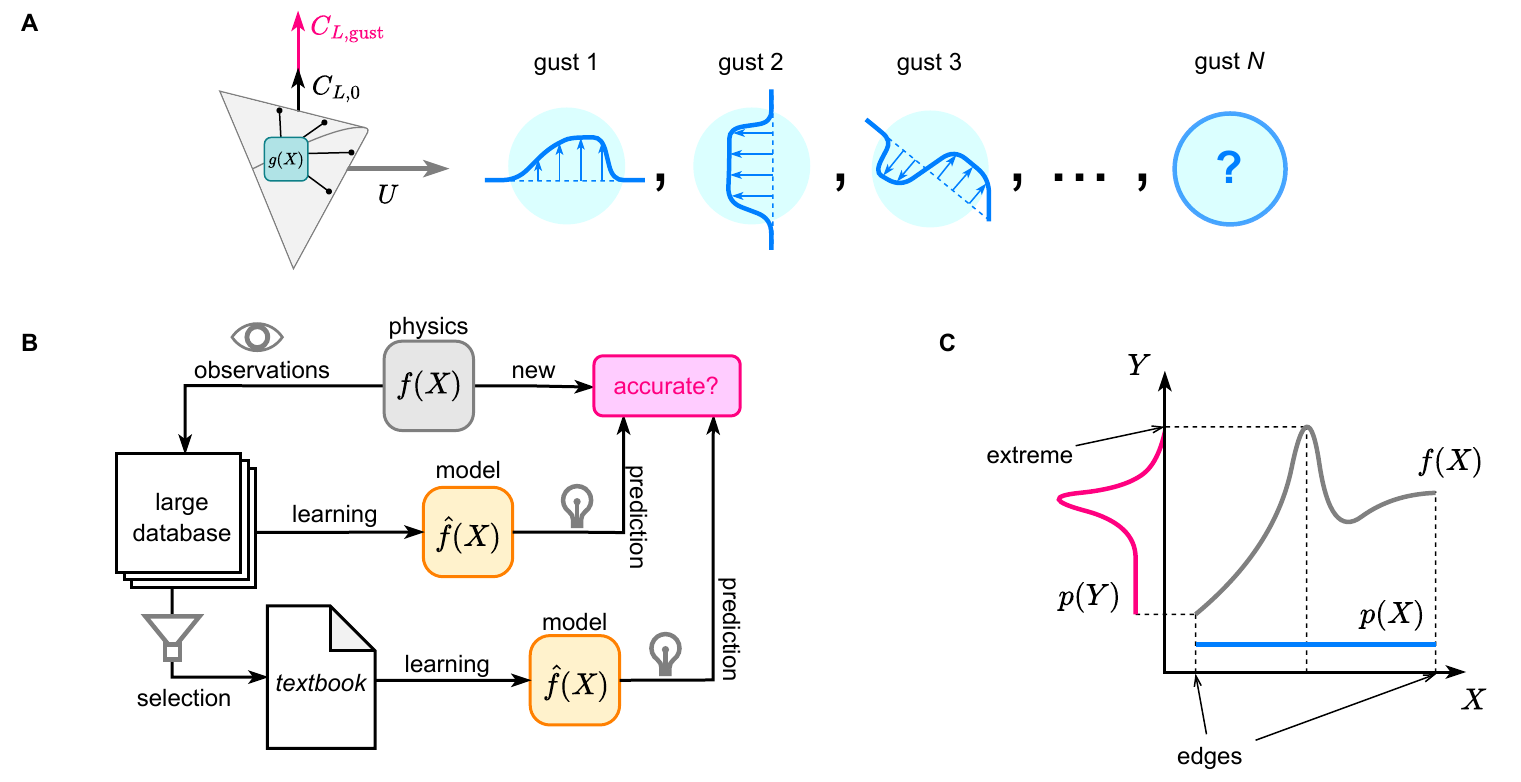}
\caption{
\textbf{ Redundant and \textbook{} data in a wing-gust interaction scenario.}
(\textbf{A}) During flight, a wing encounters a series of unsteady disturbances, resulting in
gust-induced interaction forces. 
(\textbf{B}) A model fits a large corpus of experimental data, another a concise collection of
\textbook{} cases. 
Both generalize well to the underlying physics, making accurate predictions.
(\textbf{C}) Illustration of a response relation with extreme and edge cases within a general
problem of interest. 
Colors designate inputs as blue and responses as pink.
}
\label{fig:schematic_textbook}
\end{figure*}

Figure \ref{fig:schematic_textbook}a shows a series of simplified gust encounters, where
disturbances 
impinge on a wing-shaped body and result in a surging lift coefficient $C_{L,gust}$. 
In Figure~\ref{fig:schematic_textbook}c the problem is cast in more abstract terms as a physical
system that maps an input $X$ (symbolizing the parameters describing an incoming gust) to an output
$Y$ (e.g. the unsteady aerodynamic load).  
The general aim is to learn the response function $Y=f(X)$ from data, that is from a finite number
of $X,Y$ pairs observed in experimental trials. 

Since real-world observations are typically noisy, or affected by inessential factors, it is assumed
that data are generated from a probability distribution $p(Y|X)$, which represents all that is
learnable about $f(X)$.  
We wish to collect a training database of $X,Y$ pairs of sufficient number and quality to let a
model $\widetilde{f}$ learn an accurate approximation of $f(X)$. 
As $p(Y|X)$ is initially unknown, the least biased strategy is to draw an independent
equidistributed sample of the input $X_i \sim p(X)$ of sufficient size $m_{\rm iid}$, run
experiments to measure the associated responses $Y_i  \sim p(Y|X_i)$, and compile the database
$\mathcal{D}=\{(X_i,Y_i)\}_{i=1}^{m_{\rm iid}}$.  


However, such a database is almost certainly redundant, since the geometry of $f(X)$ is complex
and likely includes features that have disproportionate relative influence on model training, such as
edge and extreme cases.  
Edge cases lie on the boundary of the physically admissible input space; extreme cases are
relatively rare inputs that lead to extreme or anomalous responses. 
While Figure~\ref{fig:schematic_textbook}c sketches a one-dimensional problem, the parameter space of
complex phenomena such as gust-wing interactions is typically very high dimensional and the data lie 
on a manifold of convoluted topology. 



\subsection{The \textbook{}}

We ask whether it is feasible to find a \textbook{} subset $\mathcal{D}_{\rm
txt}\subset\mathcal{D}$ of the data of size $m_{\rm txt}<m_{\rm iid}$, which is in some
sense an optimally representative summary of the whole input-response relation for the sake of
learning an accurate predictive model. 
We understand a \textbook{} as a subset that is optimal in that it is: i) representative,
reflecting diversity of the full-scale data, including extreme and edge cases; ii) accurate,
containing examples that are of sufficient number and quality to guarantee high predictive accuracy;
and iii) parsimonious, being be as small as possible while fulfilling the previous two properties. 
It is also desirable that the \textbook{} be reusable and general and not tied to the chosen class of
predictive model.

In data science and machine learning, the problem of dataset summarization is studied under various
guises. 
Techniques for non-redundant data subset selection exist that are either grounded solely on
geometrical or statistical similarities within the data \cite{Cormode2020,Bilmes2022} or rely
on the outcome of  a predictive model \cite{Cohn1996a,Lapedriza2013}. 
Closely connected is the concept of dataset distillation, whereby a novel, succinct dataset is
synthesized from a larger (typically generative) model trained on the full dataset \cite{Wang2018,Gunasekar2023}.  
A distinct way of handling this problem is the field of active learning where, contrary to the main
premise of our inquiry, a summary is not extracted from a large pre-existing database but
is constructed from the bottom up by gradually updating beliefs \cite{Settles2012,Ren2021,Fan2019}.  

\begin{figure}
\includegraphics[width=\linewidth]{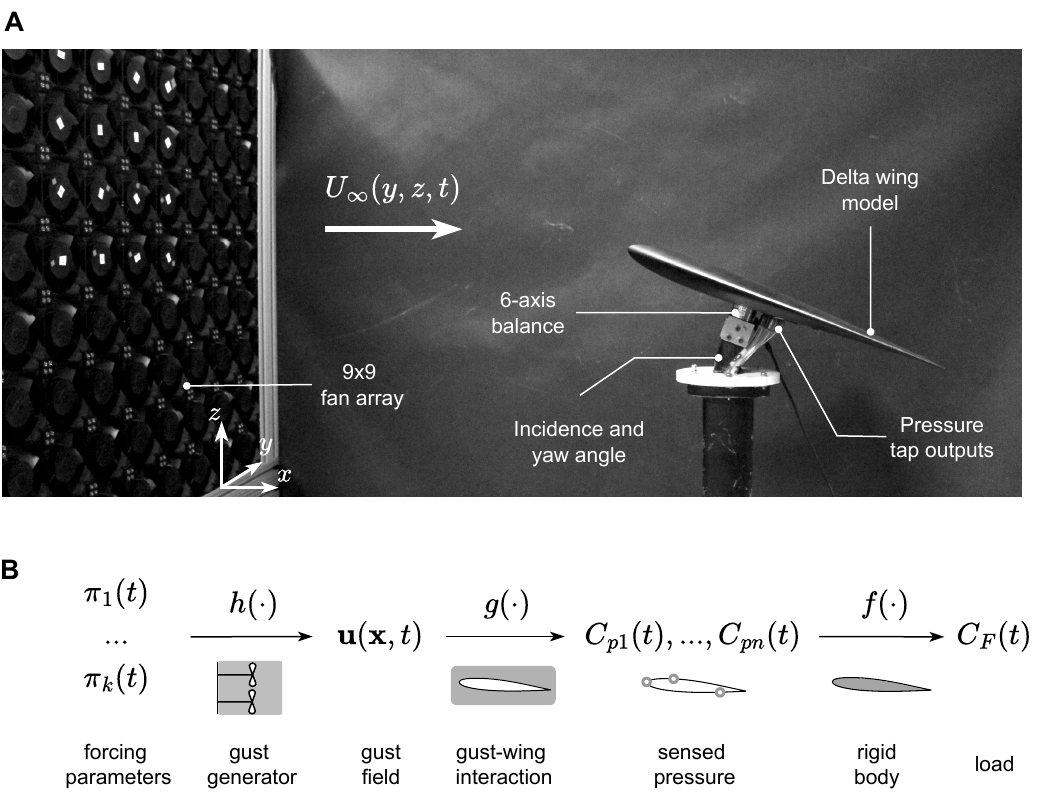}
\caption{
\textbf{Gust-wing interaction experiment.}
a) The random gust generator (seen  on the left) using a DC fan array allows us to produce axial gusts across
a broad range of conditions, subsequently felt on the non-slender delta wing mounted on its sting
(seen on the right). 
b) Graphical representation of the dependency between the relevant physical variables in a
gust-wing interaction experiment.
\label{fig:experiment}
}
\end{figure}

Here, the search for a \textbook{} will be directed towards an illustrative case of
gust-induced wing loading investigated by experimental means.
The conceptual structure of the problem is schematized by the diagram in 
Figure~\ref{fig:experiment}b. 
Arrows represent physical response operators, which from left to right are the experimental gust
generator and its fluid environment ($h$), the wing's boundary geometry and its attitude ($g$), and
the wing's geometry and inertia ($f$).
Variables include the forcing parameters $\pi_i$ of the gust generator, the resulting gust flow field
$\mathbf{u}$, the induced $n$-point discrete pressure distribution $C_{pi}$, and a response force
coefficient $C_F$. 

Within this framework, we will concentrate on learning from the data the most downstream response
$f$, i.e. the load on a wing subjected to a gust-induced, empirically measured distribution of 
pressure, and search for its \textbook{}.
In the language of the abstract model in Fig.~\ref{fig:schematic_textbook}c, $X$ is a collection of
pressure readings and $Y$ is a force coefficient acting upon the wing. 
The experimental setup and the sampling procedure are described at length in
Section~\ref{sec:experimental_database}. 

\section{Approach overview}\label{sec:experimental_database}

\subsection{Random gust generator}\label{sec:gust_generator}
The main experimental apparatus, referred to as random gust generator, is depicted in
Figure~\ref{fig:experiment}a.  
It uses a computer-controlled array of 81 dual tube-axial DC fans that allows us to produce unsteady
inflow in the form of random axial gusts. 
A generic non-slender delta wing model, with a NACA0012 cross-section, was adapted from previous
studies \cite{Marzanek2019,Burelle2020}. 
The model has a mid-span chord of $c=30$cm and incorporates internal pressure galleries, with four
pressure taps near the leading edge, whose precise location is determined based on a recent
optimization study \cite{Chen2023}.   
The wing is mounted on a support with a six-component balance, which alongside the pressure taps
collect the time-resolved data. 
Furthermore, a sting was used to vary both static angles of attack ($\alpha$) and yaw ($\beta$). 
The three forcing parameters $\pi_1, \pi_2, \pi_3$ of the gust generator are respectively the base
fan velocity, the duration of the forcing interval and the velocity increment within each forcing
interval.  
The fan array is operated at the base fan velocity, and then spun up or down to reach the value
of the velocity increment at the end of the forcing interval. 
This sequence then repeats over the next forcing interval.


\subsection{Gust-event database}\label{sec:gust_segmentation}
The gust generator is operated in a randomized fashion, to provide a dense uniform coverage of the
input parameter space.
The base fan velocity is varied among $100$ regularly spaced discrete values.
The lowest and highest base velocity values are such that the chord-based Reynolds number stays
within the range $6 \cdot 10^{4} < c U_\infty / \nu < 3.5 \cdot 10^5$ where $U_\infty$ is the
steady-state freestream speed and $\nu$ the fluid's viscosity.  
The velocity increments are explored using an equidistributed random input ranging from $10\%$ to
$130\%$ of the mean freestream speed.  
The non-dimensional forcing interval is also uniformly randomized with a minimum duration of
$G^*=2$, where $G^*=G/c$ and $G$ is defined as the gust wave length. 
All trials were carried out at the same attitude of $\alpha=30^{\circ}$, $\beta=0^{\circ}$.
The raw measurements thus consist of $100$ independent trials, each with a duration of one minute,
equating to approximately $1300$ reference convective times, returning the time-series for 4 instantaneous
pressure readings as well as the corresponding 6 aerodynamic forces and moments acting on the model.
The four pressure readings, normalized by $2 U_0^2 \rho $ to yield the time-dependent dimensionless
coefficients $C_{p0}(t)$, $C_{p1}(t)$, $C_{p2}(t)$, $C_{p3}(t)$, are the independent variables. 
The aerodynamic response variable of interest is the dimensionless lift coefficient $C_L(t) = 2 F_y / c^2 U_0^2 \rho$, with $F_y$ the vertical force measured by the balance and $U_0=6.5$m/s the reference freestream speed. 
The pool of time-resolved data is cumulatively equivalent to a flight duration of approximately $13,000$ convective times. 
The objective in selecting the data collection time frame was to create representations of typical flight envelopes. 

Each time trial then underwent a segmentation procedure to extract individual unsteady lift events.
The rationale behind segmentation is to expose the diversity of the gust ensemble that arises from the temporally randomized forcing of the fan array. 
The reader can find more on event segmentation in the Appendix.

\begin{figure}
  \centering
  \includegraphics[width=\linewidth]{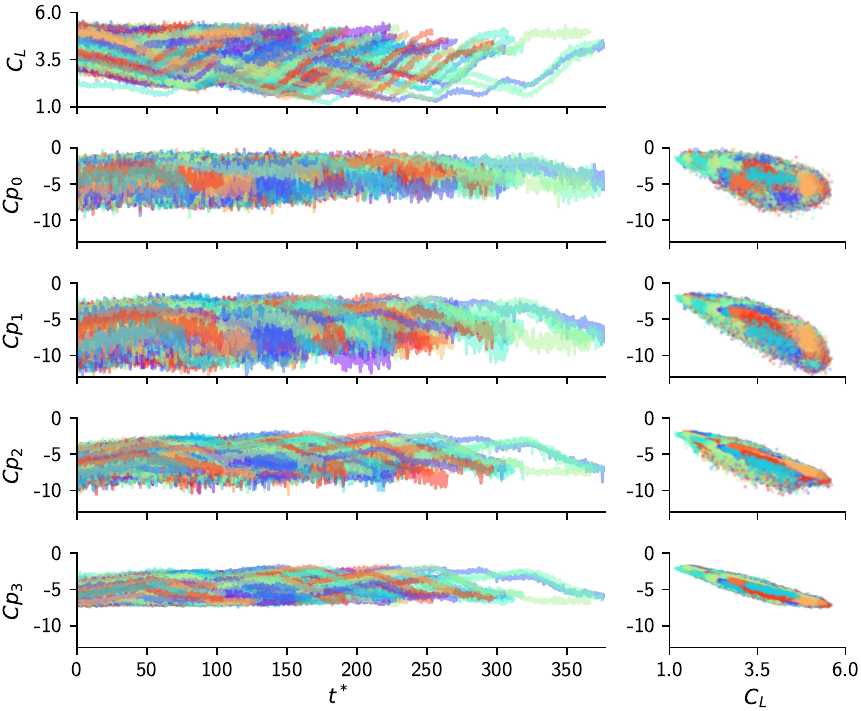}
  \caption{
  \textbf{The gust-load event dataset.}
  The left-hand side column displays the components of the five-dimensional event time-series. 
  Time is expressed in convective units as $t^*=t U_\infty / c$. 
  The right-hand column visualizes the input-output space through projections of the data on four
  coordinate planes. 
  }
  \label{fig:database_events}
\end{figure}

The final dataset is illustrated in Figure~\ref{fig:database_events} and is composed of 1031 gust
events.  
The median event duration is equivalent to approximately $110$ convective times. 


\subsection{Load prediction model}\label{sec:ml_model} 

We wish to demonstrate the concept in a basic aerodynamic load prediction setting.
The modelling task consists of predicting the instantaneous vertical load, without retaining
sequential or past information from the time-series. 
A Multi-Layer Perceptron (MLP) algorithm is adopted as the aerodynamic load prediction model. 
MLP is a basic type of feed-forward neural network that can approximate arbitrary continuous
functions \cite{Cybenko1989,Grinsztajn2022}.  
The instantaneous pressure readings from the 4 taps ($C_{p0}$, $C_{p1}$, $C_{p2}$, $C_{p3}$) serve
as the input layer of the MLP and the lift coefficient data from the load balance ($C_L$) is used as
the output.  
The full database of 1031 gusts is split $80$-$20$ into a training set and a test set of
respectively 824 and 207 events. 
The MLP is trained by minimizing the mean squared loss (MSE) on the output $C_L$.
The network architecture is optimized on the training set via hyperparameter grid search and has 4
hidden layers of 16 neurons with PReLU activation functions, amounting to a total of 977 parameters.


\section{Results}\label{sec:results} 

\subsection{Are all training data equally valuable?}\label{sec:learning_curves} 

Thanks to the predictive model and the gust database, we are now in the position to perform two
numerical tests that help draw insight into the nature of training data and thus to test the
idea of \textbook{} in a practical setting. 

\begin{figure}
\includegraphics[width=\linewidth]{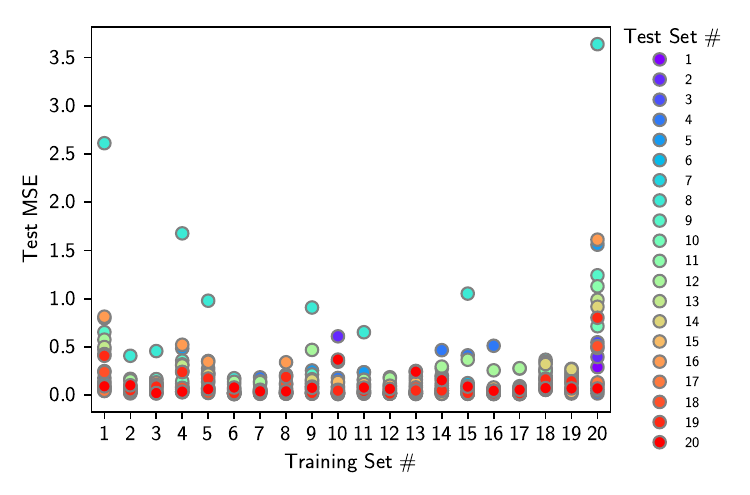}
\caption[short]{
\textbf{Training value of individual gust-load events.}
Shown are the test mean squared error (MSE) values for 29 instances of the load prediction
algorithm, each trained on a single event, identified by a number on the x-axis label,
and evaluated on another individual event, marked by different colors. 
}
\label{fig:model_mse}
\end{figure}

In the first test, a random subset of 20 gust events was picked from the main database and 20 instances of the MLP model were individually trained each on a single event, thus holding out information from the remaining 19 datasets.  
Then, the models' generalization accuracy was assessed through the MSE loss on the 19 events not encountered during training. 
The outcome is shown in Figure~\ref{fig:model_mse}.
The purpose of this procedure is to expose whether any one of the 20 training events is more (or less) informative for the sake of predicting unseen events relative to the others. 
If some training sets are indeed more (or less) informative than others, then it can be expected
that they would exhibit consistently superior (inferior) test accuracy across all test sets. 
In Figure~\ref{fig:model_mse} it is apparent that some data are informative for predicting some of
the unseen events but much less others. 
Event \#6 stands out as the best predictor across all test events. 
A handful of events are remarkably less valuable for predicting most others, such as \#20, which
results in considerable test errors. 
Most models underperform on event \#8, marked by the pale green dots that consistently make up the top error outliers.  


\begin{figure}
  \centering
  \includegraphics[width=\linewidth]{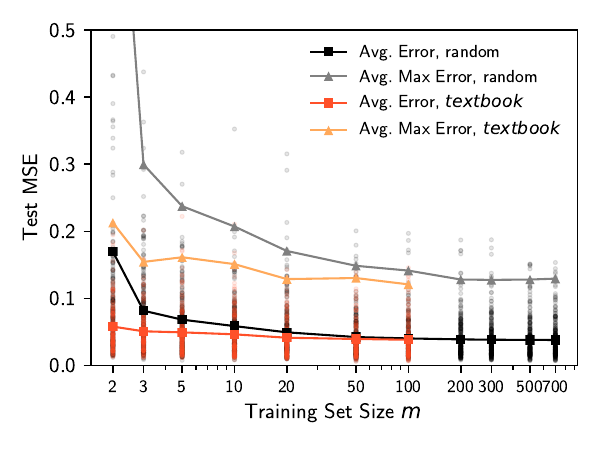}
  \caption{
  \textbf{Diminishing returns of training data.}
  The accuracy gains of a model trained on an increasing amount of data decline until learning
  completes. 
  A model trained on \textbook{} datasets will learn at a faster rate, signifying near-optimal data
  utilization. 
  The small dots are MSEs on individual events in the test set, while lines represent averaged values. 
  }
  \label{fig:learning_curves}
\end{figure}

The second numerical test gauges the effect that exposing the model to a progressively larger
amount of data has on its ability to generalize.
The model is trained on random subsets of $m$ events from the training dataset and the average
loss on the test set is evaluated. 
Figure~\ref{fig:learning_curves} depicts the outcome, the so-called learning curves of the model.
Both the expected test error and the expected maximum test error on a pool of 50 independent gusts
are shown for 11 models trained on $2$ to $700$ events, representing a coverage of
$0.2\%$ to $85\%$ of the full training set.   
On the smallest training sets, the model overfits the scarce data and is on average unable to
generalize well, analogously to what was noted in the first numerical test for models trained on
single events. 
The test error subsequently drops as the model gains access to more information until about
$m_\infty=500$ where it reaches an asymptote of minimum error and minimum uncertainty and stops
improving further, implying the model has exhausted its learning capacity above the noise level. 
This fact raises the question of whether such convergence could be accelerated by a clever choice of
high-value training examples, as implied by Figure \ref{fig:model_mse}, where some data are seen to generalize better than others.  

The challenge of compiling a \textbook{} thus amounts to assembling an optimal training dataset for the
model by selecting, for a desired \textbook{} size, the data that maximize generalization accuracy
-- or, equivalently, picking the minimal number of examples that matches a prescribed level of
accuracy.  



\subsection{\emph{Textbook} selection}\label{sec:textbook_selection}
The naive approach to compiling a \textbook{} is to train the model on all possible combinations of
examples in the base dataset and then to pick the one providing the best test accuracy.
In practice, brute-force search is computationally intractable even on moderately sized datasets.
For instance, searching for a subset of $m_{\rm txt}=5$ elements in a database of $m_{\rm iid}=1031$
examples equates to fitting the model to $\approx 10^{13}$ different combinations. 

Therefore, efficient or approximate algorithms are necessary. 
Here we take an unsupervised approach that does not rely on fitting the model to assess the value of
the data, instead using a surrogate that necessitates two elements.  
The first is a score fuction $\phi(Z)$ that, given a subset $Z\subset\mathcal{D}$ of size $m$,
measures how well the elements of $Z$ represent $\mathcal{D}$; 
the second is a procedure to maximize $\phi(Z)$ efficiently over the entire space of possible
$m$-combinations to single out the optimally representative one. 
The optimal subsets can be then verified \textit{a posteriori} by fitting the model, avoiding the
high costs incurred by repeated training. 

We choose $\phi(Z)$ to be a facility location function, a set function that constructs a
representativeness score of a subset based on a pairwise similarity metric between the data. 
Finding a subset that maximizes $\phi(Z)$ is analogous to performing combinatorial clustering of the
data, with the elements of the subset being the cluster centers.  
This choice also has the advantage that $\phi(Z)$ has the submodular property and can be thus
maximized inexpensively using the greedy heuristic \cite{Krause2014}. 
A more rigorous description of the score function and the optimization algorithm can be found in the
Supplementary Materials. 




By repeating such search procedure for several values of $m$, we obtain a series of
\textbook{}\textit{s} of increasing size. 
These are subsequently validated by training the model on each of them and recording their test
accuracy. 
The resulting learning curves are compared in Figure~\ref{fig:learning_curves} to those previously
calculated from random subsets of the data of corresponding size. 
As the curves lie below those previously computed for iid data, it follows that a \textbook{} is
always a better alternative to an average random combination of identical size. 

A \textbook{} of a specific size $m_{\rm txt}$ can be chosen based on an accuracy criterion, i.e. as the
most concise summary that guarantees an expected test accuracy to within a tolerance $e$ of that
attained in the large-sample limit. 
If we choose a tolerance of $e=20\%$ the \textbook{} has $m_{\rm txt}=10$ elements and represents a
data reduction of $98\%$ compared to the large-sample limit of around $m_{\rm iid}=500$, and is
just $1.2\%$ of the size of the full training dataset. 
A \textbook{} of $m_{\rm txt}=50$ events comes within $e=1\%$ of the maximum expected accuracy of
random data. 
It is interesting to notice that even for a \textbook{} of just $2$ events, test accuracy is about
$50\%$ worse than the large-sample limit, albeit with suboptimal worst-case accuracy.
If data are compared at constant size, a 2-event \textbook{} is $65\%$ more accurate than
the average 2-element training set and a 10-element \textbook{} is $20\%$ more accurate.

\begin{figure}
\centering
\includegraphics[width=\linewidth]{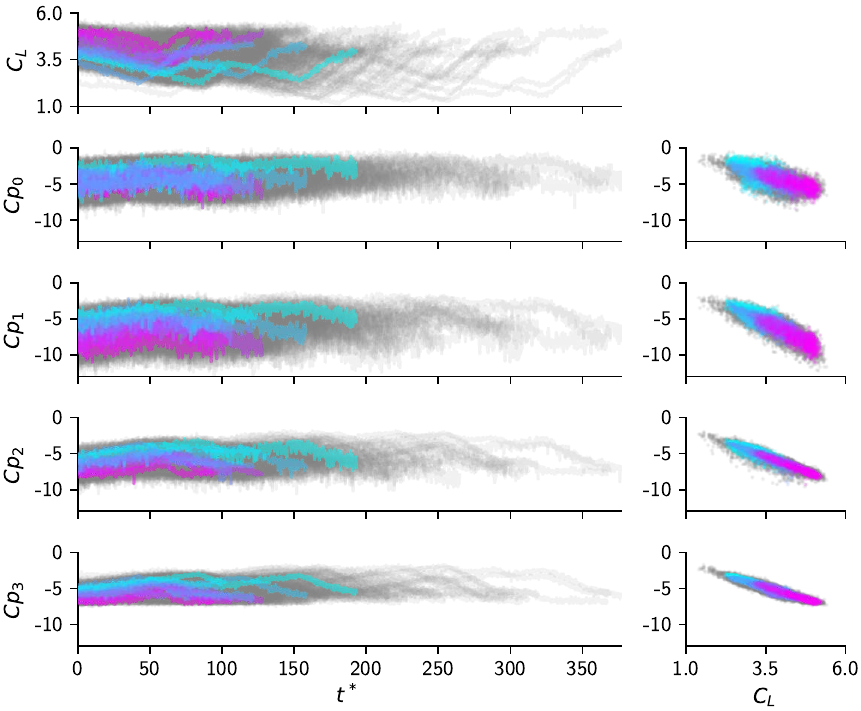}
\caption{
\textbf{Visualization of the \textbook{} dataset.}
The gust events belonging to the 10-element \textbook{} are highlighted in color and the remainder
of the base dataset is greyed out. 
Plotting conventions are as in Figure~\ref{fig:database_events}.
}
\label{fig:textbook}
\end{figure}

Figure~\ref{fig:textbook} depicts the elements of the 10-element \textbook{}, qualitatively
suggesting its role as a representative sample of the full database.
The 10 events are diverse, residing in partially non-overlapping regions of the five-dimensional
input-output space, and taken collectively represent the input-output relation well enough for the
model to learn a good approximation of the overall response succesfully.  

As an estimate of information density of a \textbook{}, we calculate the sample efficiency relative
to the full training set i.e. the ratio ${\rm MSE}_{\rm txt} m_\infty / {\rm MSE}_\infty m_{\rm
txt}$ between the fraction of the large-sample test error attained by a \textbook{} and its size
expressed as a fraction of the full training set. 
The sample efficiency of a 2-element \textbook{} is over $200$ times that of the full database,
while a 10-element \textbook{} amounts to a $50$-fold increase in sample efficiency.

If the criterion for \textbook{} selection is the worst-case expected accuracy on 50 gusts rather
than that on all loads, a similar analysis applies.  
Albeit in principle the worst-case accuracy reaches the large-sample limit later than the expected
accuracy, in the present case by $m_{\rm iid}=500$ both have been approximately achieved.
Applying the same tolerance criterion, a \textbook{} of $m_{\rm txt}=10$ is also eligible to become
the worst-case \textbook{} with similar data efficiency gains.
However, the worst-case accuracy advantage of \textbook{} data at equal $m$ is more pronounced than
on all data, reducing test error by $76\%$ for 2-event training sets and by $27\%$ for 10-eventtraining sets. 


\begin{figure*}
\includegraphics[width=\linewidth]{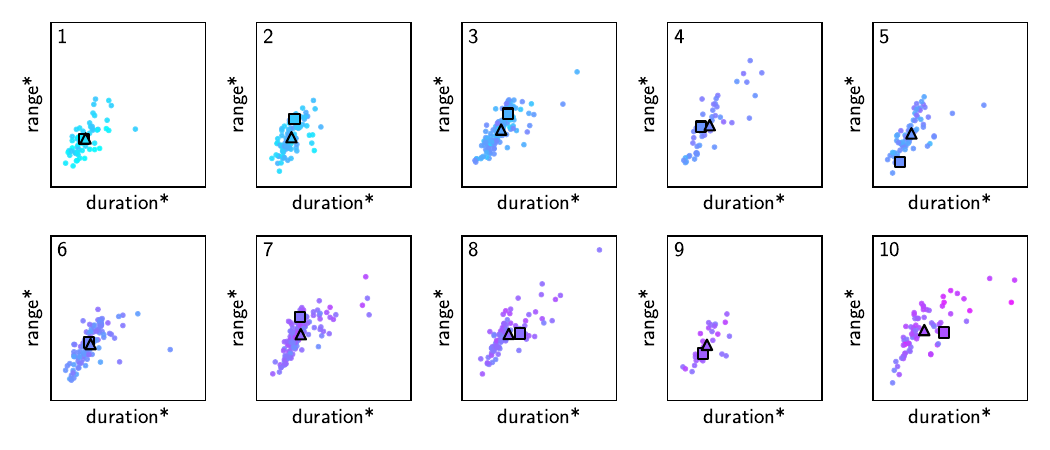}
\caption{
\textbf{Characterization of \textbook{} events.}
Each inset contains a \textbook{} event (a square marker) and the subset of events it represents (smaller circles).
The plots are sorted in ascending order of mean lift, which is also indicated for each event by its color, as in Fig.~\ref{fig:textbook}.
The asterisk marks quantities normalized to the unit interval.
The triangle markers are the barycenters of each cluster in the range-duration coordinates.
\label{fig:interpretation}
}
\end{figure*}

\subsection{Characterization of \textbook{} events}
In this section we elaborate on the question of whether a \textbook{}  reflects the diverse event types encompassed by the full database and can be used a tool to understand its diversity. 
This problem can be posed at two levels, amounting to moving from right to left up the causal chain sketched in Figure~\ref{fig:experiment}b. 
The first and lowest level concerns how the \textbook{} compares to the full dataset in the input-response space; 
the second level amounts to asking whether \textbook{} events subsume similar (spatio)temporal patterns, notwithstanding the loss of temporal information which is intrinsic in modeling the response.

The first question can be posed in terms of the geometry of the \textbook{} and the data distribution.
Figure~\ref{fig:textbook} depicts the elements of the 10-element \textbook{}, qualitatively suggesting its role as a representative sample of the full database.
The 10 events are diverse, residing in partially non-overlapping regions of the five-dimensional input-response space, and taken collectively represent the input-output relation well enough for the model to learn a good approximation of the overall response.  


To tackle the second question, we first assign each event to its closest \textbook{} event in input-output space, creating clusters.
Then, we characterize each event through their duration, the peak-to-peak range of $C_L$ and the mean $C_L$ as a basic set of time-series properties. 
Fig.~\ref{fig:interpretation} shows the outcome of this procedure.
The duration and the range are positively correlated, albeit with some outliers, and their value intervals are largely similar across most clusters, except for number 9. 
In many clusters, the square and the triangle symbols sit closely, suggesting that \textbook{} gust selected for their value in capturing the response function are also reasonably representative of some underlying time-series properties.

A separate question concerns the simulated flight conditions (the randomized one-minute trials) by which the events were generated. 
For each event we identify the trial of origin, before segmentation.
We observe that each \textbook{} event originates from a distinct one-minute trial and that clusters are heterogeneous in origin, with between 20 to 40 different flight conditions in each cluster.
The notable outlier to this trend is number 9, which is not only the smallest cluster of all but almost entirely consists of events from a single flight condition, which is also not represented in any other cluster.



\section{Summary and Discussion}\label{sec:discussion}

This study considers the question of whether an abundant experimental database on a complex
aerodynamic phenomenon can be summarized down to a significantly smaller subset -- a \textbook{}
-- that retains a sufficient amount of information to warrant high general predictive accuracy. 
We examine this problem in the context of gust-load prediction on a delta wing and we focus
on the simplified setting of aerodynamic load prediction from sparse pressure sensors.    
Key challenges include how to efficiently source a large number of experimental data and how
to effectively search such large database for the \textbook{}. 

A purpose-built automated gust generator executes a large number of randomized trials on a delta-wing model, which are then processed via a similarity-based, model-independent data summarization approach to select \textbook{}s of increasing size.  
Such summaries are then used as the training set of a machine-learning model in order to single out a \textbook{} based on the desired accuracy tolerance and its size-weighted accuracy.

We find that models trained on a \textbook{} exhibit predictive accuracy comparable to randomly
acquired training data of size up to two orders of magnitude as large, thus providing outstanding 
learning efficiency and the benefit of better interpretability. 
The resulting gust \textbook{}s are representative of the large data set in that the model generalizes more efficiently across all cases, typical and extreme, hinting that it captures the essential information of the data set.
Furthermore, we provide an example of how the elements of the \textbook{} can be characterized to inquire on the diversity of a large database of physical data in a concise manner.

We have demonstrated an approach to searching for a \textbook{} in the relatively simple setting of a gust-load prediction problem; nevertheless, the efficiency gains of data summarization are expected to grow with the complexity and the dimensionality of the problem.  
A direction we are actively pursuing is how to use the reduced number of \textbook{} examples to facilitate the extraction of common physical features of the observed phenomenon. 
One possibility is to train smaller machine-learning models on \textbook{} data that can be amenable to be more interpretable or be able to more efficiently learn optimal representations. 
Another direction is to explore how a \textbook{} could be extracted through self-supervised techniques, based on models which learn optimal low-dimensional representations of the data. 
Techniques that can blend prior physical information into the (supervised or unsupervised) summarization procedure to enhance the explanatory power of a \textbook{} are also worthy of consideration. 
Among application scenarios in unsteady fluid mechanics, the search for a \textbook{} could be particularly advantageous to problems in spatially modulated gusts and more generally to problems where time-dependent modeling is of key importance. 




\backmatter






\section*{Declarations}

We would like to thank Winston Hu and Tye Dougherty for their assistance in data collection.
D.R. gratefully acknowledges the ``Zukunft Niedersachsen'' program; 
D.R. and P.O. acknowledge the Niedersächsisches Ministerium für Wissenschaft und Kultur for financial support.

\begin{appendices}








\section{Details on data preparation} 
\begin{figure}[h!]
\includegraphics[width=\linewidth]{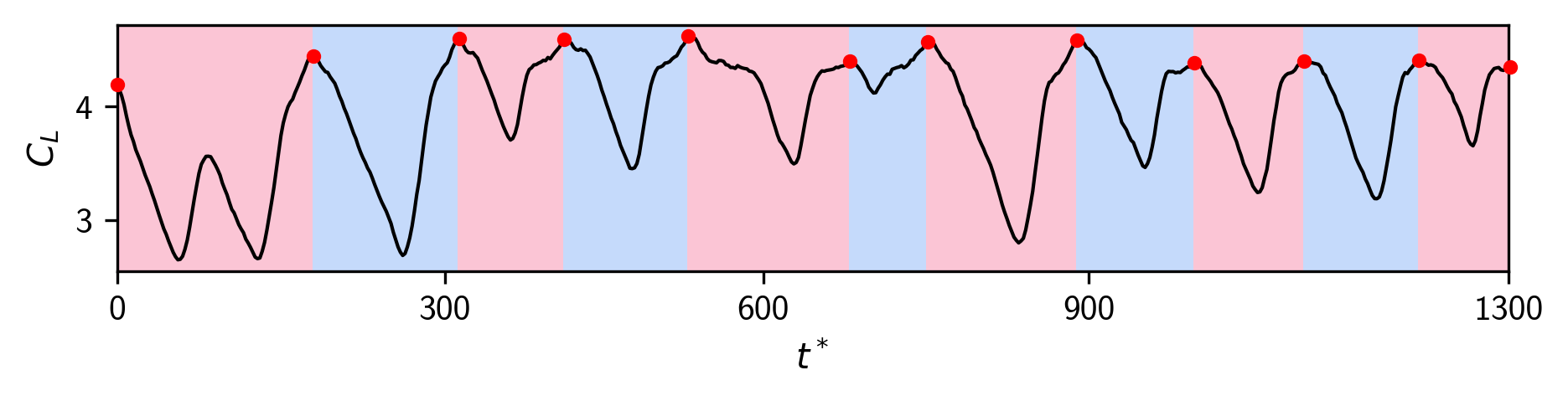}
\caption{
\textbf{Gust-load event segmentation.}
Segmentation of a typical one-minute random trial of the lift coefficient $C_L$ into gust events. 
Time is expressed in convective units as $t^*=t U_\infty / c$. 
\label{fig:gust_segmentation}
}
\end{figure}

The automated segmentation procedure consists of computing the mode of each $C_L(t)$ time series as
a proxy for the steady state, and detecting points where $C_L(t)$ reverts to the mode. 
An event is defined as the time series segment between two successive reversion points.
The time trials are pre-processed through low-pass filtering and post-processed to discard unsatisfactory events. 
The segmentation of a time trial specimen is shown in Figure~\ref{fig:gust_segmentation} , where segmentation points are marked by red dots and individual events are distinguished by background color.

\section{Details on {\textbook{}} selection} 
Compiling a \textbook{} necessitates an algorithm for searching the most representative subset of
a desired size.  
This can be broken down into adopting a quantifier of how representative a subset is of the
base dataset, and an optimization procedure to maximize such quantifier. 

We are seeking to search data summaries in an unsupervised manner, not based on training and
evaluating the model but on some easily computed geometrical or statistical property of the data. 
The representativeness of a subset $Z$ of size $|Z|=m$ with respect to database $\mathcal{D}$ is
assessed through the facility location function \cite{Mirchandani1990}, which reads   
\begin{equation}\label{eq:facility_location}
\phi(Z; \mathcal{D}) = \sum\limits_{z_j \in \mathcal{D}} \max_{z_i \in Z} s(z_i, z_j).
\end{equation}
where $s(z_i, z_j)$ is the representativeness of a single datum $z_i$ with respect to another $z_j$
which can be understood as a pairwise similarity score. 
The facility location function gives higher scores to subsets with higher average pairwise
similarities between the elements of $\mathcal{D}$ and their nearest element in $Z$, which is
analogous to a clustering criterion where the $m$ cluster centres are the elements of $Z$.  

\begin{figure}[h!]
\centering
\includegraphics[width=0.7\linewidth]{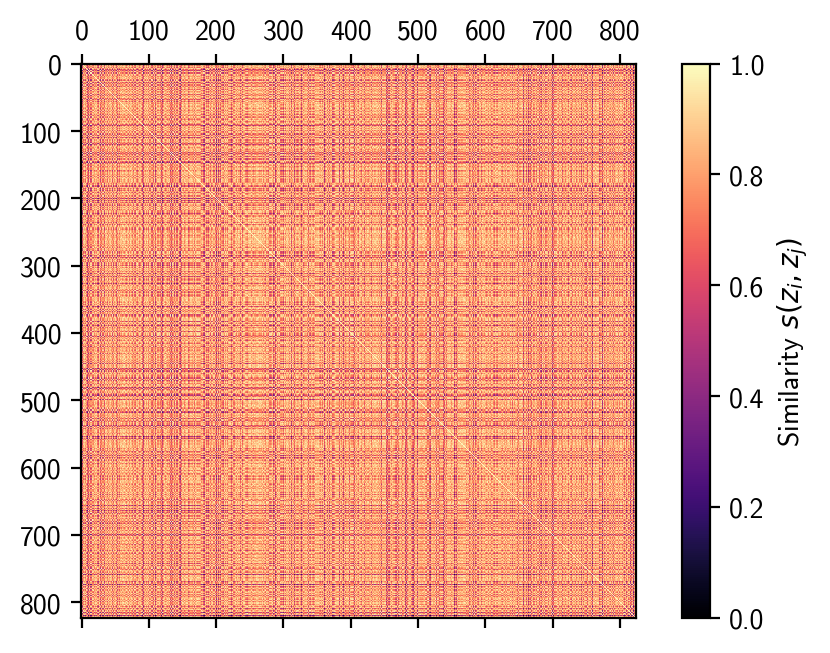}
\caption{
{\bf Pairwise similarity matrix of the training dataset.}
Each pixel of the $824 \times 824$ matrix is the similarity between the $i$-th and $j$-th element of
the training dataset.  
}
\label{fig:similarity}
\end{figure}

As our data consist of vectors of inhomogeneous length (the gust events), geometric
dissimilarity scores such as Euclidean or cosine distance can not be applied straightforwardly.  
Hence we choose a basic approach, first reducing each time-series to its $L_2$-barycenter
$\overline{z_i}$ and then calculating the Euclidean distance between the barycenters $d_{ij}=|| \overline{z_i}
- \overline{z_j} ||_2$, a computationally inexpensive operation.
We then define pairwise similarity as   
\begin{equation}\label{eq:similarity_metric}
    s_{ij} \equiv s(z_i, z_j) = \frac{\max_{ij} d_{ij} - d_{ij}}{\max_{ij} d_{ij}}.
\end{equation}
Figure \ref{fig:similarity} depicts $s_{ij}$ for the $824$ examples in the training set.
The matrix is symmetric owing to commutativity of the similarity score (\ref{eq:similarity_metric}).


A summary $Z$ of size $m$ is selected by maximizing $\phi$ over $Z\subset \mathcal{D}$,
which is a problem of combinatorial complexity. 
However, the function (\ref{eq:facility_location}) with a non-negative pairwise similarity has the
submodular property, which naturally encodes the notion of diminishing returns. 
Submodular set functions can be efficienty maximized through the greedy algorithm with proven
convergence guarantees to at least $63\%$ of the optimal value \cite{Krause2014}. 
Maximizing $\phi$ is equivalent to performing $K$-medoids clustering based on the pairwise distances
$d_{ij}$.  
The Python package Apricot \cite{Apricot2020} is the software implementation of choice of the
facility location function and the submodular optimizer.

\end{appendices}


\bibliography{dfg}

\end{document}